\documentstyle[12pt,amssymb]{article}
\begin{document}

\tolerance=5000

\def\pp{{\, \mid \hskip -1.5mm =}}
\def\cL{{\cal L}}
\def\be{\begin{equation}}
\def\ee{\end{equation}}
\def\bea{\begin{eqnarray}}
\def\eea{\end{eqnarray}}
\def\tr{{\rm tr}\, }
\def\nn{\nonumber \\}
\def\e{{\rm e}}

\begin{titlepage}

\begin{center}
\Large
{\bf Modified gravity with $\ln R$ terms and cosmic acceleration}

\vfill

\normalsize

\large{ 
Shin'ichi Nojiri$^\spadesuit$\footnote{Electronic mail: nojiri@nda.ac.jp, 
snojiri@yukawa.kyoto-u.ac.jp} and 
Sergei D. Odintsov$^{\heartsuit\clubsuit}$\footnote{Electronic mail:
 odintsov@ieec.fcr.es Also at TSPU, Tomsk, Russia}}

\normalsize

\vfill

{\em $\spadesuit$ Department of Applied Physics, 
National Defence Academy, \\
Hashirimizu Yokosuka 239-8686, JAPAN}

\ 

{\em $\heartsuit$ Institut d'Estudis Espacials de Catalunya (IEEC), \\
Edifici Nexus, Gran Capit\`a 2-4, 08034 Barcelona, SPAIN}

\ 

{\em $\clubsuit$ Instituci\`o Catalana de Recerca i Estudis 
Avan\c{c}ats (ICREA), Barcelona, SPAIN}

\end{center}

\vfill

\baselineskip=24pt
\begin{abstract}


The modified gravity with $\ln R$ or $R^{-n} \left(\ln R\right)^m$ terms which grow at small
curvature is discussed. It is shown that such a model which has 
well-defined newtonian limit may eliminate the need
for dark energy and may provide the current cosmic acceleration.
It is demonstrated that  $R^2$ terms are important
not only for early time inflation but also
to avoid the  instabilities and  the linear growth of the 
gravitational force. It is very interesting that the condition of no linear 
growth for gravitational force coincides with the one for scalar mass in the
equivalent scalar-tensor theory to be very large. Thus, modified gravity 
with $R^2$ term seems to be  viable classical theory.

\end{abstract}


\noindent
PACS numbers: 98.80.-k,04.50.+h,11.10.Kk,11.10.Wx

\end{titlepage}

\noindent 
{\bf 1. Introduction.}
The astrophysical data from high redshift surveys of type Ia
supernovae
\cite{R} and from the anisotropy power spectrum of CMB \cite{B} change 
 our image of current universe which seems to be accelerating. 
The theoretical foundation which is used to construct such universe is
also quickly evolving.
In particular, the popular explanation of current universe acceleration is
based 
on the dominance of some mysterious, exotic matter called dark
energy. There is still 
no the satisfactory theoretical explanation for the origin of this exotic
matter which should appear precisely at current epoch.

Having in mind that more astrophysical data will be available soon,
it seems the right time to search for alternative explanation of 
current cosmic speed-up. In recent papers \cite{CDTT,CCS} (see also
\cite{v1}) it has been suggested the gravitational alternative for dark 
energy. The key idea is to modify the Einstein action by $1/R$ term 
which dominates at low curvature. Remarkably, that such terms may be
predicted by some compactifications of string/M-theory \cite{sn}.
Unfortunately, it has been found \cite{Dolgov,Woodard} that $1/R$ model
contains 
unacceptable instabilities which does not appear in the Palatini version
of
the theory \cite{v,meng,fla}. (It is known that in general case 
Palatini and standard metric versions are not the same, basically they lead to different physical systems \cite{mauro}). Moreover, the  scalar-tensor 
theory which seems to be non-realistic\cite{chiba} in the standard
formulation due to
solar system observations \cite{will} becomes viable in Palatini form.

Nevertheless, more complicated modification of Einstein gravity of
the form $R^2+R+1/R$ \cite{sn1} predicts the unification of the early time
inflation and late time cosmic acceleration in the standard , metric
formulation. Moreover, the instability found in ref.\cite{Dolgov} is 
significally suppressed by higher derivative(HD) $R^2$ term or other 
HD terms like $R^3$. The solar system test for
equivalent scalar-tensor theory \cite{chiba} may be passed 
because  scalar has large mass induced again by HD terms. 
The consideration of such theory in Palatini form has been recently done
\cite{wang}. It is shown that account of $R^2$ term makes the theory
viable also in Palatini form.

In the present paper we continue the search for the realistic modified
gravity which 
may provide the gravitational alternative for dark energy. As such a
model,
we suggest  to account for the $\ln R$ terms in modified gravity.
Such terms are basically induced by quantum effects in curved spacetime. 
Various versions of such modified gravity may eliminate the
need for dark energy and may serve
for the unification of the early time inflation and cosmic acceleration.
HD terms again suppress the instability and improve the 
solar system bounds so that the theory may be viable. The correction to the
gravitational coupling constant 
may be small enough too.

\noindent
{\bf 2. General formulation and simplified model.}
One may start from rather arbitrary function $F(a,b_\mu,c,g_{\mu\nu})$
which depends on 
two scalars $a$, $c$, one vector $b_\mu$ and metric $g_{\mu\nu}$. 
The general starting action is: 
\be
\label{RD1}
S={1 \over \kappa^2}\int d^4 x \sqrt{-g} F(R, \partial_\mu R, \square R, g_{\mu\nu})\ .
\ee
Here $R$ is the scalar curvature. Introducing the auxiliary fields $A$ and $B$, 
one may rewrite the action (\ref{RD1}) as following:
\be
\label{RD2}
S={1 \over \kappa^2}\int d^4 x \sqrt{-g} \left\{B\left(R-A\right) 
+ F(A, \partial_\mu A, \square A, g_{\mu\nu})\right\}\ .
\ee
By the variation over $B$,  $A=R$ follows. Substituting it into (\ref{RD2}), 
the action (\ref{RD1}) can be reproduced. Making the variation with
respect to $A$ first, we obtain
\bea
\label{RD3}
B=f(A)&\equiv& \left.\left(\partial_a F(a, b_\mu, c, g_{\mu\nu})
 - \nabla_\mu \left(\partial_{b_\mu} F(a, b_\mu, c, g_{\mu\nu})\right) \right.\right. \nn
&& \left.\left.+ \Box \left(\partial_c F(a,b_\mu,c,g_{\mu\nu})\right)\right)
\right|_{a=A, b_\mu = \partial_A, c=\square A}\ ,
\eea
which may be solved with respect to $A$ as $A=G(B)$. 
In general, $A$ is solved non-locally as a function of $B$. 
Eliminating $A$ in (\ref{RD2}) by using $G(B)$, we obtain
\be
\label{RD5}
S={1 \over \kappa^2}\int d^4 x \sqrt{-g} \left\{B\left(R-G(B)\right) 
+ F\left(G(B), \partial_\mu G(B), \square G(B),g_{\mu\nu}\right)\right\}\ .
\ee
Instead of $A$, one may eliminate $B$  and arrive at the
equivalent action. 

The scalar field $\sigma$ may be defined $\sigma = - \ln B = \ln f(A)$. 
One can scale the metric by $g_{\mu\nu}\to \e^\sigma g_{\mu\nu}$. 
Then the action (\ref{RD5}) can be rewritten as 
\bea
\label{RD9}
S&=&{1 \over \kappa^2}\int d^4 x \sqrt{-g} \left\{ R - {3 \over 2}g^{\mu\nu}
\partial_\mu \sigma \partial_\nu \sigma - V(\sigma)\right\} \nn
V(\sigma)&\equiv&  G\left(\e^{-\sigma}\right) \e^\sigma - \e^{2\sigma}
F\left(G\left(\e^{-\sigma}\right), \partial_\mu \left(G\left(\e^{-\sigma}\right)\right),
\right.\nn && \qquad \left. \square \left(G\left(\e^{-\sigma}\right)\right) 
+ g^{\mu\nu}\partial_\mu \sigma 
\partial_\nu \left(G\left(\e^{-\sigma}\right)\right), \e^\sigma g_{\mu\nu}\right)\ .
\eea
It is given in the Einstein frame. On the other hand, the
(physical) action 
(\ref{RD5}) is given in the Jordan frame. 

As the simple example, we consider the following case:
\be
\label{RD10}
F(A)=A + \alpha \ln {\square A \over \mu^4}\ .
\ee
One finds
\be
\label{RD11}
\e^{-\sigma}=B= 1 + \alpha \square \left({1 \over \square A}\right)\ ,
\ee
which can be solved non-locally with respect to $A$.
Then the complicated expression for the potential follows
\bea
\label{RD13}
&& V(\sigma) =  \e^\sigma \square^{-1}\left({\alpha \over \square^{-1} 
\left(\e^{-\sigma} - 1\right)}\right) - \e^{2\sigma}\left[
\square^{-1}\left({\alpha \over \square^{-1} \left(\e^{-\sigma} - 1\right)}\right) \right.\\
&& \left. + \alpha \ln \left\{{\alpha \over \mu^4 \square^{-1} \left(\e^{-\sigma} - 1\right)} 
+ g^{\mu\nu}\partial_\mu \sigma \partial_\nu \left(
\square^{-1}\left({\alpha \over \mu^4\square^{-1} 
\left(\e^{-\sigma} - 1\right)}\right) \right)\right\}\right]\ .\nonumber
\eea

We may consider more general example
\be
\label{RD14}
F(A)=A + \alpha \ln {\square A \over \mu^4} + \alpha' \ln {A \over \mu^2} 
+ \beta A^m\ .
\ee
This model is very complicated. If we consider the case $R$ is almost
constant, 
the second term turns to the (cosmological) constant.
The natural starting model  looks as follows:
\be
\label{RD15}
F(A)=A + \alpha' \ln {A \over \mu^2} + \beta A^m\ .
\ee
Furthermore,  $m=2$ choice  simplifies the model.
The generalizations of this model will be considered at final section.
The correspondent $R^2$ term at large curvature leads to well-known trace
anomaly driven (Starobinsky) inflation. 
Assuming the scalar curvature is constant and the Ricci tensor is also 
covariantly constant, the equations of motion corresponding to the action (\ref{RD1}) 
are:\footnote{
If  the action contains only $R^2$ term, $F(R)=\beta R^2$, 
Eq.(\ref{RD19}) is trivial, which means the arbitrary constant curvature
 space 
 is a solution. Adding the Einstein-Hilbert
term, 
$F(R)=R + \beta R^2$, the only solution of (\ref{RD19}) is $R=0$.
 Then if we start  
with large $R$ solution, which may correspond to the inflation, due to the
Einstein-Hilbert term, 
$R$ decreases slowly and goes to zero. The inflation will
be stopped.
}
\be
\label{RD19}
0=2F(R) - RF'(R) = f(R) \equiv R + 2\alpha' \ln {R \over \mu^2} - \alpha'\ .
\ee
If $\alpha'>0$, $f(R)$ is monotonically increasing function and we have
$\lim_{R\to 0}f(R)=-\infty$ and $\lim_{R\to +\infty}f(R)=+\infty$. There is 
one and only one solution of (\ref{RD19}). This solution may correspond 
to the inflation. On the other hand, if $\alpha'<0$, 
 $\lim_{R\to 0}f(R)=\infty$ and $\lim_{R\to +\infty}=+\infty$. Since 
$f'(R) = 1 + {\alpha' \over R}$, the minimum of $f(R)$, where $f'(R)=0$, is 
given by $R=-2\alpha'$. If $f(-2\alpha')>0$, there is no solution of
(\ref{RD19}). 
If $f(-2\alpha')=0$, there is only one solution and if $f(-2\alpha')<0$, 
there are two solutions\footnote{It is interesting that only $\ln 
R$ may lead to early time inflation and current acceleration. The
AdS cosmologies \cite{cvetic} may be considered as well.}. Since
$f(-2\alpha')=-2\alpha'\left(1 - \ln {-2\alpha' \over \mu^2}
\right)$, there are two solutions if $- {2\alpha' \over \mu^2}>\e$. 
Since the square root of the curvature $R$ corresponds to the rate of the expansion 
of the universe, the larger solution in two solutions might correspond to the 
inflation in the early universe and the smaller one to the present accelerating 
universe. 

For $m=2$ case, we have 
\be
\label{RD16}
\e^{-\sigma} = 1 + {\alpha' \over A} + 2\beta A\ ,
\ee
which can be solved with respect to $A$: 
\be
\label{RD17}
A={-\left(1 - \e^{-\sigma}\right)\pm \sqrt{\left(1 - \e^{-\sigma}\right)^2 - 8\beta\alpha'} 
\over 4\beta}\ .
\ee
In the branch that the r.h.s. in (\ref{RD16}) is negative, one may choose 
\be
\label{RD17b}
\e^{-\sigma} = -\left( 1 + {\alpha' \over A} + 2\beta A\right)\ ,
\ee
Then instead of (\ref{RD9}), we obtain 
\bea
\label{RD9b}
S&=&{1 \over \kappa^2}\int d^4 x \sqrt{-g} \left\{ - R + {3 \over 2}g^{\mu\nu}
\partial_\mu \sigma \partial_\nu \sigma - \tilde V(\sigma)\right\} \nn
\tilde V(\sigma)&\equiv&  -G\left(\e^{-\sigma}\right) \e^\sigma - \e^{2\sigma}
F\left(G\left(\e^{-\sigma}\right)\right)\ .
\eea
As the sign in front of the scalar curvature $R$ is changed, this seems
to be the indication to
the anti-gravity. Of course, the anti-gravity should not be real 
but apparent since the physical 
theory in the real spacetime should be given by (\ref{RD1}) with
(\ref{RD15}) ($m=2$ case). 

The potential is
\be
\label{RD18}
V(A)=\tilde V(A) 
= { \alpha'\left(1 - \ln {A \over \mu^2}\right) + \beta A^2 \over 
\left( 1 + {\alpha' \over A} + 2\beta A \right)^2} \ .
\ee
Then in terms of $\sigma$, $V(\sigma)$ can be expressed as
\bea
\label{RD21}
V(\sigma)&=& \e^{2\sigma}\left\{ \alpha' \left( 1 - \ln {-\left(1 - \e^{-\sigma}\right)\pm 
\sqrt{\left(1 - \e^{-\sigma}\right)^2 - 8\beta\alpha'} \over 4\beta \mu^2}\right) \right. \nn
&& \left. + \beta \left( {-\left(1 - \e^{-\sigma}\right)\pm 
\sqrt{\left(1 - \e^{-\sigma}\right)^2 - 8\beta\alpha'} \over 4\beta}\right)^2 \right\}\ .
\eea
 $\tilde V(A)$ can be expressed in a similar way.  
If $A$ is small, from (\ref{RD16}) it follows $A\sim \pm
\alpha'\e^\sigma$. Here $+$ ($-$) 
sign corresponds to the case that the r.h.s. in (\ref{RD16}) is positive (negative), 
which also corresponds to the case that $\alpha'>0$ ($\alpha'<0$). 
Then $A\to 0$ corresponds to $\sigma\to -\infty$ and $V(\sigma)$ ( $\tilde
V(\sigma)$ )
behaves as 
\be
\label{RD22}
V(\sigma) = \tilde V(\sigma) \sim - {A^2 \over \alpha'}\ln {A \over \mu^2}\ .
\ee
or 
\be
\label{RD22c}
V(\sigma) = -\tilde V(\sigma) \sim -\alpha' \e^{2\sigma}\sigma\ .
\ee
On the other hand, when $A$ is large, we find
\be
\label{RDD1}
V(A)\to {1 \over 4\beta}\left(1 - {2 \over \beta A}\right)\ .
\ee
Then $V(A)$ is monotonically increasing function for large $A$ if $\beta>0$ 
and approaches to a constant ${1 \over 4\beta}$. 
In order to find the extrema of $V(A)$ and $\tilde V(A)$ one 
differentiate them 
with respect to $A$:
\be
\label{RDD2}
V'(A)=\tilde V'(A) = { \left( - {\alpha' \over A^2} + 2\beta\right)
\left(A - \alpha' + 2\alpha' \ln {A \over \mu^2}\right) \over 
\left( 1 + {\alpha' \over A} + 2\beta A \right)^3} \ .
\ee
 $V'(A)=0$ if $A - \alpha' + 2\alpha' \ln {A \over \mu^2}=0$, which 
corresponds to (\ref{RD19}). Furthermore if $\alpha', \beta>0$ there appears a new extremum when 
$- {\alpha' \over A^2} + 2\beta=0$ or $A=A_0\equiv \sqrt{\alpha' \over 2\beta}$. 
As $A_0$ is not a solution of (\ref{RD19}), this may be artificially 
caused by the rescaling. In fact, $A_0$ corresponds to the point that the mapping 
(\ref{RD16}) is degenerate, that is, ${d\e^{-\sigma} \over dA}=0$ at $A=A_0$. 
Anyway we may discuss the (in)stability of the solution in (\ref{RD19}) by using 
the potential $V(A)$. When $\alpha', \beta>0$, let the solution corresponding 
to (\ref{RD19})
 be $A_1$. Combining (\ref{RD22}), (\ref{RDD1}), and (\ref{RDD2}), 
one finds that if $A_0<A_1$ ($A_0>A_1$), the solution of
(\ref{RD19}) ($A=A_1$) 
is locally stable (instable). 
On the other hand, if $\alpha'<0$ and $\beta>0$, there is a singularity in 
$V(A)$ when $1 + {\alpha' \over A} + 2\beta A=0$.  
 It corresponds to $\sigma\to +\infty$. 
Let the two solutions (if exist) in (\ref{RD19})  be $A_\pm$
($A_-<A_+$). 
Combining (\ref{RD22}), (\ref{RDD1}), and (\ref{RDD2}) again, 
we find that there are three cases: 1. When $A_2<A_-<A_+$, $A_+$ ($A_-$) is 
locally stable (instable). 2. When $A_-<A_2<A_+$, both of $A_+$ and $A_-$ are 
locally stable. 3. When $A_-<A_+<A_2$, $A_-$ ($A_+$) is 
locally stable (instable).

\begin{figure}
\begin{center}
\unitlength=0.8mm
\begin{picture}(160,100)
\thicklines
\put(20,20){\vector(1,0){120}}
\put(20,20){\vector(0,1){60}}
\qbezier[100](20,20)(30,20)(35,50)
\qbezier[100](35,50)(40,80)(50,80)
\qbezier[100](50,80)(55,80)(60,75)
\qbezier[100](60,75)(65,70)(75,70)
\qbezier[100](75,70)(85,70)(95,75)
\qbezier[150](95,75)(105,80)(140,80)
\put(15,82){$V(A)$}
\put(142,18){$A=R$}
\put(130,82){\circle*{4}}
\put(128,82){\vector(-1,0){10}}
\put(75,72){\circle*{4}}
\put(73,13){$R_0$}
\put(128,13){$R_{\rm initial}$}

\thinlines
\put(73,72){\vector(-1,0){5}}
\put(77,72){\vector(1,0){5}}
\qbezier[40](130,79)(130,40)(130,20)
\qbezier[35](75,70)(75,40)(75,20)

\end{picture}
\end{center}
\caption{ A typical potential when $\alpha'>0$. We may start with large curvature 
$A=R=R_{\rm initial}$ (inflation). Then the curvature rolls down the potential slowly 
and stops at the small curvature $A=R=R_0$ (the present accelerating unvierse). 
\label{Fig1}}
\end{figure}
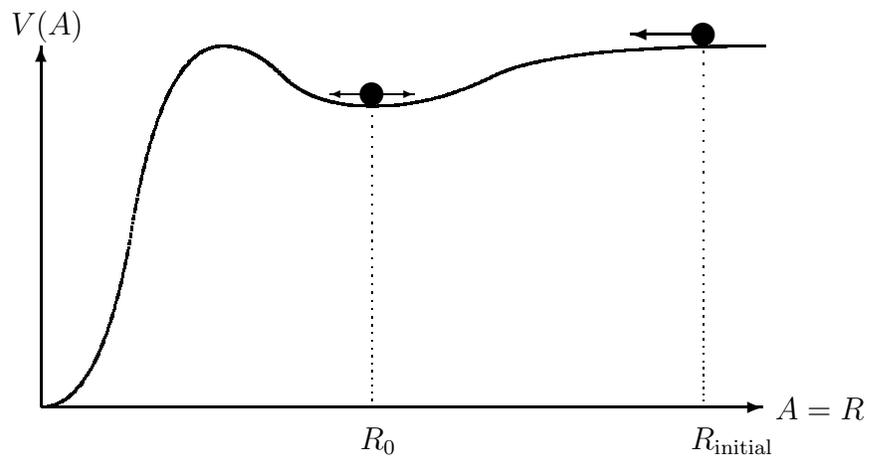

When $\alpha'>0$, a typical (conceptual) potential is given in Figure \ref{Fig1}. 
There may be the following scenario of the inflation and of the
present accelerating 
universe: If we start with large curvature $A=R=R_{\rm initial}$, the inflation 
occurs due to the large curvature. Since the potential is slowly
increasing, the 
curvature rolls slowly down the potential and becomes smaller. If the 
curvature reaches  the local minimum $A=R=R_0$, the curvature will stay there 
with non-trivial value $R=R_0\neq 0$, which may correspond to the present 
acceleration of the universe. As the potential plays a role of the cosmological 
constant, the ratio $w$ of the effective pressure with respect to the effective 
energy density becomes minus unity, $w=-1$. As we will show later in (\ref{SS5}), 
by fine-tuning the parameters, the $\sigma$-field can become heavy and decouple.

\noindent
{\bf 3. Cosmic deceleration in $\ln R$ gravity.} 
Let us assume the metric in the physical (Jordan) frame (\ref{RD5}) is 
given in the FRW form:
\be
\label{RD23}
ds^2 = - dt^2 + \hat a(t)^2 \sum_{i,j=1}^3 \hat g_{ij} dx^i dx^j\ .
\ee
In the Einstein frame, the FRW equation looks like
\be
\label{RD24}
3H_E^2 = {3 \over 4}{\dot\sigma}^2 + {1 \over 2}V(\sigma)\ \mbox{or}\  
3H_E^2 = {3 \over 4}{\dot\sigma}^2 - {1 \over 2}\tilde V(\sigma)\ .
\ee
Here we distinguish the quantities in the Einstein frame by the subscript $E$. 
The Hubble parameter $H_E$ is now defined by $H_E\equiv {\dot {\hat a}_E \over {\hat a}_E}$. 
On the other hand, the equation derived by the variation over $\sigma$ is 
\be
\label{RD25}
0=3\left(\ddot\sigma + 3H_E\dot\sigma\right) + V'(\sigma)\ \mbox{or}\ 
0=3\left(\ddot\sigma + 3H_E\dot\sigma\right) - \tilde V'(\sigma)\ .
\ee
We now consider the case that $A$ or the scalar curvature in the Jordan 
(physical) frame is small and the potential is given by (\ref{RD22}). 
Then the  solution (in the leading order of $t_E$) of the combined
equations (\ref{RD24}) and (\ref{RD25}) is given by 
\be
\label{RD26}
\sigma \sim -\ln {t_E \over t_0} + {1 \over 2}\ln \ln {t_E \over t_0}\cdots
\ ,\quad a_E \propto t_E^{3 \over 4} + \cdots\ .
\ee
The time coordinate $t_E$ in the Euclidean frame is related with 
the time coordinate $t$ in the (physical) Jordan frame by $\e^{\sigma \over 2}dt_E 
= dt$. As a result $t_0^{1 \over 2}t_E^{1 \over 2}\propto t$. 
The power law inflation occurs in the physical (Jordan) frame 
\be
\label{RR45}
\hat a = \e^{{\sigma \over 2}} \hat{a_E} \propto t_E^{1 \over 4} \propto t^{1 \over 2} \ .
\ee
Since we have $\dot{\hat a}>0$ but $\ddot{\hat a}<0$, the deccelerated expansion occurs. 
In case of the original $1/R$ model in \cite{CDTT}, the solution is $\hat a \propto 
t^2$. This might suggest that for the model containing 
$R^n$ $(-1<n<0)$, one may have more moderately accelerating
universe: $\hat a \propto t^m$ 
(${1 \over 2}<m<2$). 

 From (\ref{RD24}) and (\ref{RD25}), we may evaluate
\be
\label{RR45bb}
\alpha' t_0^2 \sim {\cal O}(1)\ .
\ee
Eqs.(\ref{RR45}) indicate $H ={\dot {\hat a} \over \hat a}={1 \over 2 t}$. 
Since $H^2={\kappa^2 \over 6}\rho$, the energy density $\rho$ corresponding to 
$\sigma$ may be defined as $\rho={\rho_0 \over t^2}$. 
Here $\rho_0$ is a constant. Denoting the pressure of $\sigma$ by $p$
and substituting the above expressions of $H$ and $\rho$ to the 
conservation law for the energy-momentum tensor
one finds
 $w\equiv {p \over \rho}= {1 \over 3}$, which is nothing but 
that of the radiation. 

One may account for the matter contribution to the energy-momentum tensor. 
When it is dominant compared with the one from $\sigma$, 
the obtained results are not changed from those in \cite{CDTT} where the
possibility of cosmic acceleration in $1/R$ model was established. 

\noindent
{\bf 4. Instabilities and corrections to gravitational coupling constant.}
We now discuss the (in)stability of our model under the perturbations. 
In \cite{Dolgov}, small gravitational object like the Earth or the Sun in the 
model \cite{CDTT} is considered. It has been shown that the system quickly
 becomes instable. 

Following to the idea in ref.\cite{sn1} we start from the action
(\ref{RD1}), where $F$ is given 
by (\ref{RD15}) with $m=2$. 
If  the Ricci tensor is not covariantly constant, the trace of equation of
motion with matter is given by\footnote{We should note that the convention of 
the spacetime signature here is different from those in \cite{Dolgov}.}
\be
\label{RR57}
\Box R + {F^{(3)}(R) \over F^{(2)}(R)}\nabla_\rho R \nabla^\rho R 
+ {F'(R) R \over 3F^{(2)}(R)} - {2F(R) \over 3 F^{(2)}(R)} 
= {\kappa^2 \over 6F^{(2)}(R)}T\ .
\ee
Here $T=T_\rho^{\ \rho}$. In case of the Einstein gravity, where $\alpha'=\beta=0$, the 
solution of Eq.(\ref{RR57}) is given by $R=R_0\equiv -{\kappa^2 \over 2}T$. 
The perturbation around the solution may be addressed\footnote{Due to the
spherical symmetry and the structure of the lagrangian  it is enough to
consider only perturbation of the curvature. Equivalently, 
one can transform the analysis to Jordan (Brans-Dicke) theory.
There are no problems with instabilities in the Einstein part,
while resolution of the instabilities for sigma field which corresponds to
curvature leads to 
the same bounds as in below analysis.} 

\be
\label{RR59}
R=R_0 + R_1\ ,\quad \left(\left|R_1\right|\ll \left|R_0\right|\right)\ .
\ee
Then by linearizing (\ref{RR57}), we obtain 
\bea
\label{RR60}
0&=&\Box R_0 + {F^{(3)}(R_0) \over F^{(2)}(R_0)}\nabla_\rho R_0 \nabla^\rho R_0 
+ {F'(R_0) R_0 \over 3F^{(2)}(R_0)} - {2F(R_0) \over 3 F^{(2)}(R_0)} 
 - {R_0 \over 3F^{(2)}(R_0)} \nn
&& + \Box R_1 + 2{F^{(3)}(R_0) \over F^{(2)}(R_0)}\nabla_\rho R_0 \nabla^\rho R_1 
+ U(R_0) R_1\ .
\eea
Here
\bea
\label{RR61}
&& U(R_0)\equiv \left({F^{(4)}(R_0) \over F^{(2)}(R_0)} - {F^{(3)}(R_0)^2 
\over F^{(2)}(R_0)^2}\right) \nabla_\rho R_0 \nabla^\rho R_0 + {1 \over 3}R_0 \\
&& - {F^{(1)}(R_0) F^{(3)}(R_0) R_0 \over 3 F^{(2)}(R_0)^2} 
 - {F^{(1)}(R_0) \over F^{(2)}(R_0)} + {2 F(R_0) F^{(3)}(R_0) \over 3 F^{(2)}(R_0)^2} 
 - {R_0 F^{(3)}(R_0) \over F^{(2)}(R_0)^2} \ .\nonumber
\eea
If $U(R_0)$ is negative, the perturbation $R_1$ grows up exponentially with time. 
The system becomes instable. By including the $R^2$-term, the time for instability to 
occur is significally improved  (by the order of $10^{29}$) \cite{sn1}, compared with 
the original $1/R$ model in \cite{CDTT}. If the coefficient of $R^2$ is
already fixed by some other condition, one can use  other HD
terms (like $R^3$) to eliminate the instability completely.  

In (\ref{RR57}), we only considered the Ricci scalar. More general equation of 
motion, including all the metric components  or Ricci
tensor, has the 
following form: 
\bea
\label{RDG1}
&& {1 \over 2}g_{\mu\nu} F(R) - R_{\mu\nu} F(R) - g_{\mu\nu}\nabla_\rho\nabla^\rho
\left(F'(R)\right) \nn
&& + \nabla_\mu \nabla_\nu \left(F'(R)\right) = - {\kappa^2 \over 2}T_{\mu\nu}\ .
\eea
Let  the perturbation of the metric be $\delta g_{\mu\nu}$. We may choose
a gauge 
condition $\nabla^\mu \delta g_{\mu\nu}=0$. Furthermore we decompose $\delta g_{\mu\nu}$ 
into the sum of the trace part $\delta G=g^{\mu\nu}\delta g$ and traceless part 
$\delta \hat g_{\mu\nu}$ as 
\be
\label{A1}
\delta g_{\mu\nu} = \delta \hat g_{\mu\nu} + {1 \over 4}g_{\mu\nu}\delta G\ .
\ee
For perturbation from the background with constant curvature 
$\left(R_{\mu\nu}={3 \over l^2}g_{\mu\nu}\right)$, one gets
\be
\label{A2}
\delta R = R_1 = - {3 \over l^2}\delta G - \nabla^2 \delta G\ .
\ee
Then if $R_1$ is given, $\delta G$ is uniquely determined as in the usual
 Einstein gravity 
up to the homogeneous part $\delta G_h$ which satisfies 
$- {3 \over l^2}\delta G_h - \nabla^2 \delta G_h=0$. For the Ricci tensor $R_{\mu\nu}$, 
we have 
\be
\label{A3}
\delta R_{\mu\nu} = {2 \over l^2}\delta\hat g_{\mu\nu}
 - {1 \over 2}\nabla^2 \delta \hat g_{\mu\nu} - {1 \over 2l^2}g_{\mu\nu}\delta G 
 - {1 \over 8}g_{\mu\nu}\nabla^2 \delta G - {1 \over 2}\nabla_\mu \nabla_\nu \delta G\ .
\ee
Therefore if we use (\ref{RDG1}), the traceless part
 $\delta \hat g_{\mu\nu}$ can be also 
uniquely determined up to the homogeneous part $\delta \hat g_{h\,\mu\nu}$
which satisfies ${2 \over l^2}\delta\hat g_{h\,\mu\nu}
 - {1 \over 2}\nabla^2 \delta \hat g_{h\,\mu\nu}=0$, 
as in the usual Einstein gravity. 
Since the homogeneous parts $\delta G_h$ and $\delta \hat g_{h\,\mu\nu}$ 
appear in 
the usual perturbation from the background of the deSitter space,
 they are not related with the 
(in)stability. Then if $\delta R=R_1$ is stable, whole the metric perturbation 
$\delta g_{\mu\nu}$ is also stable. 

In \cite{Woodard}, it has been found that the linearly growing force appears in $1/R$ 
model due to a diffuse source in a locally deSitter background:
\be
\label{WR1}
ds^2 = -dt^2 + \e^{2Ht} \sum_{i=1,2,3}\left(dx^i\right)^2\ ,
\ee
which is a solution, with a constant curvature $R_0=12 H^2$, of the equation corresponding to 
(\ref{RR57}) or (\ref{RD19}) for the vacuum case. 
If we consider the perturbation  
 (\ref{RR59}) by assuming $R_1=R_1(y)$ ($y\equiv \e^{Ht} H \sqrt{\sum_{i=1,2,3}\left(x^i
\right)^2}$), the solution $R_1$ is a sum of two independent solutions $f_0(y)$ and 
$f_{-1}(y)$: $R_1(y) = \beta_1 f_0(y) + \beta_2 f_{-1}(y)$. 
First, we consider 
the case that $\beta\gg \alpha' R_0^2$. Then  $U(R_0)={R_0 \over 3}$ and 
\be
\label{WR2}
f_0(y)=1 - {2 \over 3}y + {\cal O}\left(y^2\right)\ ,\quad 
f_{-1}(y)= {1 \over y}\left( 1 - 3 y + {\cal O}\left(y^2\right)\right)\ ,
\ee
for the vacuum solution. If we assume there is a spherical matter source with mass $M$ and the 
radius $r_0$ as in \cite{Woodard}, the coefficients $\beta_1$ and $\beta_2$ are determined 
by
\bea
\label{WR3}
\beta_1&=& {3MG \over r_0^3}\left\{ f_0(y_0) - {f'_0(y_0) f_{-1}(y_0) \over f_0'(y_0)}
\right\}^{-1}={3MG \over r_0^3}\left\{ 1 + {\cal O}\left(y_0^2\right)\right\}\ ,\nn
\beta_2&=& {3MG \over r_0^3}
\left\{ f_{-1}(y_0) - {f'_{-1}(y_0) f_0(y_0) \over f_{-1}'(y_0)}
\right\}^{-1}=-{4MG y_0^3 \over r_0^3}\left\{ 1 + {\cal O}\left(y_0^2\right)\right\}\ .
\eea
Here $16\pi G =\kappa^2$. It is assumed that the source exists in $y\leq
y_0$. 
Since for the size of galaxies,  $y=10^{-6}$ and for the typical distance 
between galaxies, $y=10^{-4}$, one may assume $y_0\ll y \ll 1$ \cite{Woodard}. Then 
 $\beta_2\ll \beta_1$ and term with $\beta_2$ may be neglected. 
If we denote the trace part of the perturbation of the metric by $h$, we find
\be
\label{WR4}
h'(y)=-{2 \over y^2\left(1-y^2\right)^{3 \over 2}}\int_0^y dy' {y'}^2 \left(
1 - {y'}^2\right)^{1 \over 2}{R_1(y') \over H^2} 
\sim - {2GM \over H^2 r_0^3}y + {\cal O}\left(y^3\right)\ .
\ee
Then there appears a linear growth as in \cite{Woodard}, which might be a
phenomenological disaster. However, that was the case with large 
 $\beta$. In more general case, the equation corresponding to 
(\ref{RR60}) has the following form:
\bea
\label{WR5}
0&=&\left[\left(1-y^2\right){d^2 \over dy^2} + {2 \over y}\left(1 - 2 y^2\right)
{d \over dy} + {12 U(R_0) \over R_0}\right]R_1(y)\ , \\
U(R_0)&=& { - {8 R_0^2 \beta^2 \over 3}- {4\alpha' \beta \over 3R_0} - 2\beta 
+ {{\alpha'}^2 \over 3R_0^3}\left(2 + 4\ln {R_0 \over \mu^2}\right) 
 - {\alpha' \over 3R_0^2} \over \left(-{\alpha' \over R_0^2} + 2 \beta\right)^2}\ .
\eea
As clear from Eq.(\ref{RD19}),  $R_0$ does not depend on $\beta$. 
Choosing $\beta\to {\alpha' \over 2R_0^2}$, 
$U(R_0)$ becomes very large and we find $R_1\to 0$ in the vacuum, which is 
identical with the case of the Einstein gravity with cosmological constant. 
Then contrary to the case of ref.\cite{Woodard}, there does not appear
the linear growth of $h'(y)$ but 
$h'(y)$ behaves as $y^{-2}$, which does  not conflict with the present
cosmology. 
We should also note that the condition $\beta\to {\alpha' \over 2R_0^2}$ is identical with the condition that 
$\sigma$-field decouples (\ref{SS5}), as will be shown below.
Hence, modified gravity with terms growing at small curvature and with
higher derivative terms important for early time inflation may be viable
theory.

It has been mentioned in ref.\cite{chiba} that $1/R$ model which is
equivalent to some scalar-tensor gravity is ruled out as realistic theory 
due to the constraints to such theories.
As the coupling of $\sigma$ with matter is not small \cite{fla}, we now 
calculate the square of scalar mass, which is proportional to
$V''(\sigma)$. 
We consider the fluctuation from the solution (\ref{RD26}). 
Then 
\be
\label{SS}
V''(\sigma) \sim {\alpha' t_0^2 \over t_E^2} \sim {t_0^2 \over t^4}\ .
\ee
Here we have used $t_0^{1 \over 2}t_E^{1 \over 2}\propto t$ and (\ref{RR45bb}) are used. The result 
(\ref{SS}) itself 
does not change from the original $1/R$ model \cite{CDTT}.  
Since the 
Hubble parameter is given by $H={\dot a \over a}={1 \over 2t}$, 
in the present universe, we have ${1 \over t} \sim H_0$. 
Here $H_0\sim 10^{-33}\mbox{eV}$ is the Hubble parameter of 
the present universe. Then 
\be
\label{SS3}
V''(\sigma) \sim t_0^2 H_0^4\ .
\ee
Surely $H_0$ is very small but we have no restriction (or we have not found it) 
on $t_0$. Then if $t_0$ is very large, the mass of $\sigma$ can be large. 
Assuming the mass is larger than 1 TeV, we have $t_0\sim
10^{78}\rm{eV}^{-1}$.\footnote{
The parameter $t_0$ should be determined from the initial condition 
but since there 
is unknown parameter $\alpha'$, which may differ from $\left(10^{-33}{\rm
eV}
\right)^2$ as we have argued, $t_0$ contains the ambiguity coming 
from $\alpha'$. 
As the curvature is small, 
from 
(\ref{RD16}), we have $\sigma\sim \ln {A \over \alpha'}$.
 On the other hand from 
(\ref{RD26}),  $\sigma \sim - \ln {t_E \over t_0} \sim - \ln {t \over t_0} 
\sim \ln \left(H_0^2 t_0^2 \right)$. As $A$ corresponds to the real curvature, 
one may assume $A\sim H_0^2$. Then we obtain $\alpha' t_0^2 \sim {\cal
O}(1)$, 
which is identical with (\ref{RR45bb}). In order 
to determine the value of the parameter $\alpha'$, one should use the
information related with the 
inflation and the matter contents.} 
As $\alpha't_0^2\sim {\cal O}(1)$  (\ref{RR45bb}),  
$\alpha' \sim \left(10^{-78}\rm{eV}\right)^2$. This indicates that 
such class of theories may still pass the solar system bounds for
scalar-tensor gravity. Moreover, the account of the terms with derivatives of the
curvature \footnote{The correspondent scalar-tensor theory becomes higher 
derivative one as it follows from second section.} 
may permit to pass the solar system tests even easier.

In \cite{chiba}, the PPN (Parametrized Post-Newtonian) parameters have been 
investigated in the Jordan (Brans-Dicke) frame and it has been found that 
the VLBI parameter is above the constraint due the solar system experiments. 
The analysis relied on the mass of the Brans-Dicke scalar, which corresponds 
to $\sigma$ or $A$ in this paper. In the Brans-Dicke form, we find $\omega=0$, 
that is, there is no kinetic term for $\sigma$. In the current limit, 
$\omega>3500$ \cite{will}. 
Then the Brans-Dicke scalar should be heavy to avoid the problem.  
In the Minkowski background, it has been found that the mass 
is too light and unnatural. 
In our case, the action correponding to (\ref{RD15}) can be rewritten as
\be
\label{BD1}
S={1 \over \kappa^2}\int d^4 x \sqrt{-g}\left[ 
\left(1 + {\alpha' \over A} + 2\beta A\right)R
 + \alpha' \left(\ln {A \over \mu^2} - 1\right) - \beta A^2\right]\ ,
\ee
which is also the Brans-Dicke form with $\omega=0$. 
We should note, however, the present universe is not exactly Minkowski 
but accelerating. The mass of the BD scalar is given by (\ref{SS3}). Then as discussed, 
the mass is determined by the parameter of the integration $t_0$, which may be 
determined dynamically. Such a mass may be quite large 
and then such theory does not violate solar system test as in
\cite{chiba}.

One may consider the case that the present universe corresponds to the 
solution  (\ref{RD19}). In such a case, by tuning the parameters 
$\beta$ and $\alpha'$, the mass of $\sigma$ can be made 
large again. Let us write  the solution(s) of (\ref{RD19}) as $A_1$
for 
$\alpha'>0$ case and $A_\pm$ for $\alpha'<0$ case. Then 
\bea
\label{SS4}
\left.{d^2 V(\sigma) \over d\sigma^2}\right|_{A=A_1,A_\pm}&=&
\left\{\left({d\sigma \over dA}\right)^{-2}
\left.{d^2 V(A) \over dA^2}\right\}\right|_{A=A_1,A_\pm} \nn
&=&\left.{\left(1 + {2\alpha' \over A}\right) \over \left(1 + {\alpha' \over A}
+ 2\beta A\right)\left( - {\alpha' \over A^2} + 2\beta \right)}
\right|_{A=A_1,A_\pm} \ .
\eea
Choosing
\be
\label{SS5}
\beta \sim \left.{\alpha' \over 2A^2}\right|_{A=A_1,A_\pm}\ ,
\ee
for $\alpha'>0$  or 
\be
\label{SS6}
\beta \sim - \left.\left( {1 \over 2A} + {\alpha' \over 2A^2}\right)
\right|_{A=A_1,A_\pm}\ ,
\ee
for $\alpha'<0$ , the mass of $\sigma$ becomes large again. Thus, HD term
may help to pass the solar system tests for $\ln R$ (or $1/R$) gravity. 

\noindent
{\bf 5. Discussion.}
Finally, we calculate the corrections to gravitational coupling constant.
The easiest way is to consider the perturbation around the constant
curvature solution  (\ref{RD19}). 
We now write the metric as $g^{(0)}_{\mu\nu}$ corresponding the solution and the constant 
scalar curvature as $R^{(0)}={12 \over l^2}$. 
The metric is splitted into the background part $g^{(0)}_{\mu\nu}$ and the 
perturbation $h_{\mu\nu}$:
\be
\label{RD28}
g_{\mu\nu}=g^{(0)}_{\mu\nu} + h_{\mu\nu}\ .
\ee
The following gauge conditions are imposed:
\be
\label{RD29}
0=g^{(0)\mu\nu}h_{\mu\nu}=\nabla^{(0)\mu}h_{\mu\nu}\ .
\ee
The first condition is chosen to simplify our discussion as
graviton is spin 2 field. 
 Then
\bea
\label{RD30}
R&=&R^{(0)}+ {1 \over l^2}h_{\mu\nu}h^{\mu\nu} + {1 \over 4}h^{\mu\nu}\square^{(0)}
h_{\mu\nu} + \square \left(h^{\mu\nu} h_{\mu\nu}\right) + {\cal O}\left(h^3\right)\ ,\nn
\eea
Using the equation $0=R^{(0)} + 2\alpha'\ln {R^{(0)} \over \mu^2} -
\alpha'$,  
the expanded action $\tilde S$ has the following form:
\be
\label{RD31}
\tilde S ={1 \over \kappa^2}\left(1 + {\alpha' l^2 \over 12} + {24\beta \over l^2}\right)
\int d^4x \sqrt{-g^{(0)}}\left({6 \over l^2} - {1 \over 2l^2}h_{\mu\nu}h^{\mu\nu} 
+ {1 \over 4}h^{\mu\nu} \square^{(0)} h_{\mu\nu}\right)\ .
\ee
Here  the total derivative term is dropped. In case of the Einstein action
with a 
positive cosmological constant
\be
\label{RD32}
S_E={1 \over \kappa^2}\int \sqrt{-g}\left(R - {6 \over l^2}\right)\ ,
\ee
which has  deSitter solution, the corresponding linearized action 
is:
\be
\label{RD33}
\tilde S_E ={1 \over \kappa^2}
\int d^4x \sqrt{-g^{(0)}}\left({6 \over l^2} - {1 \over 2l^2}h_{\mu\nu}h^{\mu\nu} 
+ {1 \over 4}h^{\mu\nu}\square^{(0)} h_{\mu\nu}\right)\ .
\ee
Comparing (\ref{RD33}) with (\ref{RD31}),  the gravitational
constant 
$\kappa$ is renormalized as
\be
\label{RD34}
{1 \over \kappa^2}\to {1 \over \kappa^2}\left(1 + {\alpha' l^2 \over 12} 
+ {24\beta \over l^2}\right)\ .
\ee
The Newton potential will be  modified respectively.
When $R_0={12 \over l^2}$ corresponds to the rate
of the 
expansion of the present universe, one gets
$R_0\sim {12 \over l^2}\sim \mu^2\sim 10^{-33} \mbox{eV}$. 
In (\ref{RD34}),  $\beta$-dependent term 
 is of the same order term as second term when scalar has large mass.
With the assumption $\alpha'\sim \left(10^{-33} \mbox{eV}\right)$  
the correction may be significant. We may, however, consider the case
$\alpha'$ is 
much smaller than $10^{-33}$eV, since there seems to be no constraint for the 
value of $\alpha'$ itself. We only have a constraint (\ref{RR45bb}) and $t_0$ 
can be very large. In such a case, the correction to the Newton constant
can be very small and the starting theory has acceptable newtonian limit.

\noindent

One may discuss further generalizations of modified gravity like
\be
\label{GR1}
F(A)=A + \gamma A^{-n}\left(\ln {A \over \mu^2}\right)^m\ .
\ee
Here we restrict $n$ by $n>-1$ ($m$ is arbitrary) in order that  
the second term \footnote{The sum of such terms (whose coefficients could
be constrained by the condition of avoiding the linear growing
gravitational
force) may be considered.
 The presence of $R^2$ at large curvature is supposed.} could
be more dominant than the Einstein term 
 when $A$ is small. 
 $n$ and $m$ can be fractional or irrational numbers 
in general. 

When the physical scalar curvature $A=R$ is small, we find
\bea
\label{GR2}
&& \e^{-\sigma}\sim - \gamma n A^{-n-1}\left(\ln {A \over \mu^2}\right)^m\ ,\quad 
A\sim \left( - \gamma n \e^\sigma\right)^{1 \over n+1}\left({\sigma \over n+1}
\right)^{m \over n+1}\ ,\nn
&& V(\sigma)\sim \left(1 + {1 \over n}\right) \left( -\gamma n\right)^{1 \over n+1}
\e^{{n+2 \over n+1}\sigma}\left({\sigma \over n+1}\right)^{-m}\ ,
\eea
when $n\neq 0$ and 
\bea
\label{GR3}
&& \e^{-\sigma}\sim {m\gamma \over A}\left(\ln {A \over \mu^2}\right)^{m-1}\ ,\quad 
A\sim m\gamma \e^\sigma \sigma^{m-1}\ ,\nn
&& V(\sigma)\sim \gamma \e^{2\sigma} \sigma^m \ ,
\eea
when $n=0$. With the similar procedure as in the previous sections, 
\bea
\label{GR4} 
& \sigma \sim - {2(n+1) \over (n+2)}\ln {t_E \over t_0} 
+ {m(n+1) \over n+2}\ln\ln {t_E \over t_0}+ \cdots\quad & (n\neq 0)\ ,\nn
& \sigma \sim - \ln {t_E \over t_0} 
 - {m \over 2}\ln\ln {t_E \over t_0}+ \cdots\quad & (n= 0)\ ,\nn
& {\gamma \over t_0^2} \sim {\cal O}(1)\ . 
\eea
As a result
\be
\label{GR5}
t\sim t_E^{1 \over n+2}\ ,\quad a_E \sim t_E^{3(n+1)^2 \over (n+2)^2}\ ,
a\sim t^{(n+1)(2n+1) \over n+2}\ .
\ee
This  does not depend on $m$. The logarithmic factor is 
almost irrelevant. We also find the effective $w$ for the $\sigma$-field 
is 
\be
\label{GR6}
w=-{6n^2 + 7n - 1 \over 3(n+1)(2n+1)}\ .
\ee
Then $w$ can be negative if 
\be
\label{GR7}
-1<n<-{1 \over 2}\ \mbox{or}\ 
n>{-7 + \sqrt{73} \over 12}=0.1287\cdots \ .
\ee
 From (\ref{GR5}), the condition that the universe 
could accelerate is ${(n+1)(2n+1) \over n+2}>1$, that is:
\be
\label{GR8}
n> {-1 + \sqrt{3} \over 2}=0.366\cdots \ .
\ee 
Clearly, the effective dark energy $w$ may be within the existing bounds.

Thus, we demonstrated that modified gravity with $\ln R$ or $R^{-n} (\ln R)^m$ 
terms may be responsible for the current acceleration of the universe.
Hence, like the simplest $1/R$ modified gravity this provides the
gravitational alternative for dark energy.
Moreover, the presence of HD terms like $R^2$ (which may be responsible 
for early time inflation) helps to pass the existing 
arguments (instabilities, solar system tests) against such modification of
the Einstein gravity. The theory may also have the well-acceptable
newtonian limit. It is clear that much more work is requiered to
(dis)prove that one of the versions of such modified gravity is currently
realistic theory. Nevertheless, the fine-tuning of parameters of modified
gravity to provide the effective gravitational dark energy looks more
promising than the introduction by hands some mysterious fluid.

\noindent
{\bf Acknowledgments} 
The research is supported in part by the Ministry of
Education, Science, Sports and Culture of Japan under the grant n.13135208
(S.N.),  RFBR grant 03-01-00105
(S.D.O.) and LRSS grant 1252.2003.2 (S.D.O.). 
S.N. is indebted to all the members of IEEC, especially to E. Elizalde, 
for the hospitality during the time when this work was started.

\newpage

\end{document}